\documentclass[12pt]{iopart}
\pdfoutput=1
\usepackage{geometry}
\usepackage{graphicx}
\usepackage{color}
\usepackage{units}
\usepackage{amssymb}

\definecolor{green4}{rgb}{0.0,0.7,0.0}

\newcommand{\nn}{\nonumber}
\newcommand{\bra}{\langle}

\newcommand{\ket}{\rangle}

\begin{document}
\title{Arrays of waveguide-coupled optical cavities that interact strongly with atoms}
\author{G. Lepert$^{1}$, M. Trupke$^{1}$, M.J. Hartmann$^{2}$,
M.B. Plenio$^{3,4}$, E.A. Hinds$^{1}$}
\address{$^{1}$Centre for Cold Matter, Imperial College, Prince
Consort Road, London SW7 2BW, United Kingdom}
\address{$^{2}$Technische Universit\"at M\"unchen, Physik Department,
James-Franck-Strasse, 85748 Garching, Germany}
\address{$^{3}$Universit\"at Ulm, Institut f\"ur Theoretische Physik,
Albert-Einstein-Allee 11, 89069 Ulm, Germany}
\address{$^{4}$Quantum Optics and Laser Science Group, Imperial
College London, London SW7 2PG, United Kingdom}
\ead{guillaume.lepert07@imperial.ac.uk}

\begin{abstract}

We describe a realistic scheme for coupling atoms or other quantum emitters with an array of coupled optical cavities.
We consider open Fabry-Perot microcavities coupled to
the emitters. Our central innovation is to connect the
microcavities to waveguide resonators, which are in turn evanescently coupled to each other on a
photonic chip to form a coupled cavity chain. In this paper, we describe the components, their technical limitations and the factors
that need to be determined experimentally. This provides the basis for a detailed theoretical analysis of two possible
experiments to realize quantum squeezing and controlled quantum dynamics. We close with an outline of more advanced applications.
\end{abstract}

\maketitle

\section{Introduction}

The interaction of light and matter is of central importance to research in
quantum optics~\cite{WMbook}. The strength of this interaction depends on the intensity
of the light at the location of the atom. To enhance the interaction,
photons can be trapped in a high-finesse cavity of
small volume leading to high intensity fields inside the cavity even
with small photon numbers. This arrangement forms the central paradigm
of cavity quantum electrodynamics (CQED)~\cite{WVEB06,RBH01,MD02}. Early
experiments in CQED used atoms that where dropped
through the cavity and hence passed briefly through the region of strong interaction~\cite{TRK92}. With the advent of laser cooling and trapping
techniques, it became possible to position
atoms or ions at a desired location within the cavity and to achieve ``strong'' atom-photon coupling, i.e. a coupling rate faster than both the atomic decay rate and the cavity decay rate. Phenomena such as vacuum Rabi splitting
\cite{BMB+04}, photon blockade \cite{BBM+05,Day08} and single photon
sources \cite{HWS+07} have now been observed.

So far, experiments in CQED have almost exclusively been carried
out with single cavities. In recent years however, it has become possible
to fabricate microcavities in regular arrays in various settings
\cite{AKS+03,BSP+06,AAS+03,Hen07,Wal04,THE+05,AV04,SNAA05}. Due to
their small size, these cavities all have small mode volume and correspondingly
a strong interaction between light and matter. Indeed
the ``strong coupling'' regime has already been demonstrated for
several of these devices~\cite{ADW+06,TGD+07,Hen07}.

An important next step is to find a way to couple these cavities so that they form an array in which photons can tunnel from one cavity to another.
Recently, various approaches have been put forward for using such arrays of coupled cavities
as a platform for quantum simulators~\cite{HBP06,HBP08,ASB07,GTCH06,Na08}. These include a scheme for simulating the
Bose-Hubbard Hamiltonian \cite{HBP06,HP07,HBP07b,HBP07c}, models of interacting
Jaynes-Cummings Hamiltonians \cite{ASB07,GTCH06}, and an
effective spin Hamiltonian \cite{HBP07a}. The phase diagrams of these models have been studied \cite{RF07,AHTL08,KL09,RFS08} and the existence of a glassy phase has been predicted \cite{RF07}.

In this work, we describe the design for a practical device, in which atoms interact strongly with high finesse microcavities and, at the same time, photons can tunnel with low loss from one cavity to the next through an interconnecting waveguide chip. The main components of the device are shown schematically in Fig.~\ref{fig:chip_outline}.

The paper is organised as follows. In section \ref{sec:device}, we describe the design of the device in detail and specify some realistic parameters. In section \ref{sec:spectrum} we calculate the
spectrum of one composite cavity. We then show (section \ref{sec:JC}) that the dynamics of the device is well approximated by a Jaynes-Cummings Hamiltonian. We describe experiments on a small scale to demonstrate two types
of quantum simulator using coupled cavity arrays. As a first example (section \ref{sec:JCspectroscopy}) we
show that a lattice of interacting Jaynes-Cummings Hamiltonians could
be implemented in our device and analyse its steady states in a driven
dissipative regime for a two-cavity setup \cite{Hartmann10,Leib10,Knap10,Ferretti10}.
In a second example (section \ref{sec:effspin}), we show
that our device is also suitable for implementing effective spin
Hamiltonians and we study the dynamics in a two-cavity setup.
We discuss briefly a variety of applications that can be
based on these capabilities but go beyond the two cavity setting. Finally we summarize our results and
give an outlook on future perspectives in section \ref{sec:outlook}.

\section{Device \label{sec:device}}

\begin{figure}[h]
\centering
\includegraphics[width=\textwidth]{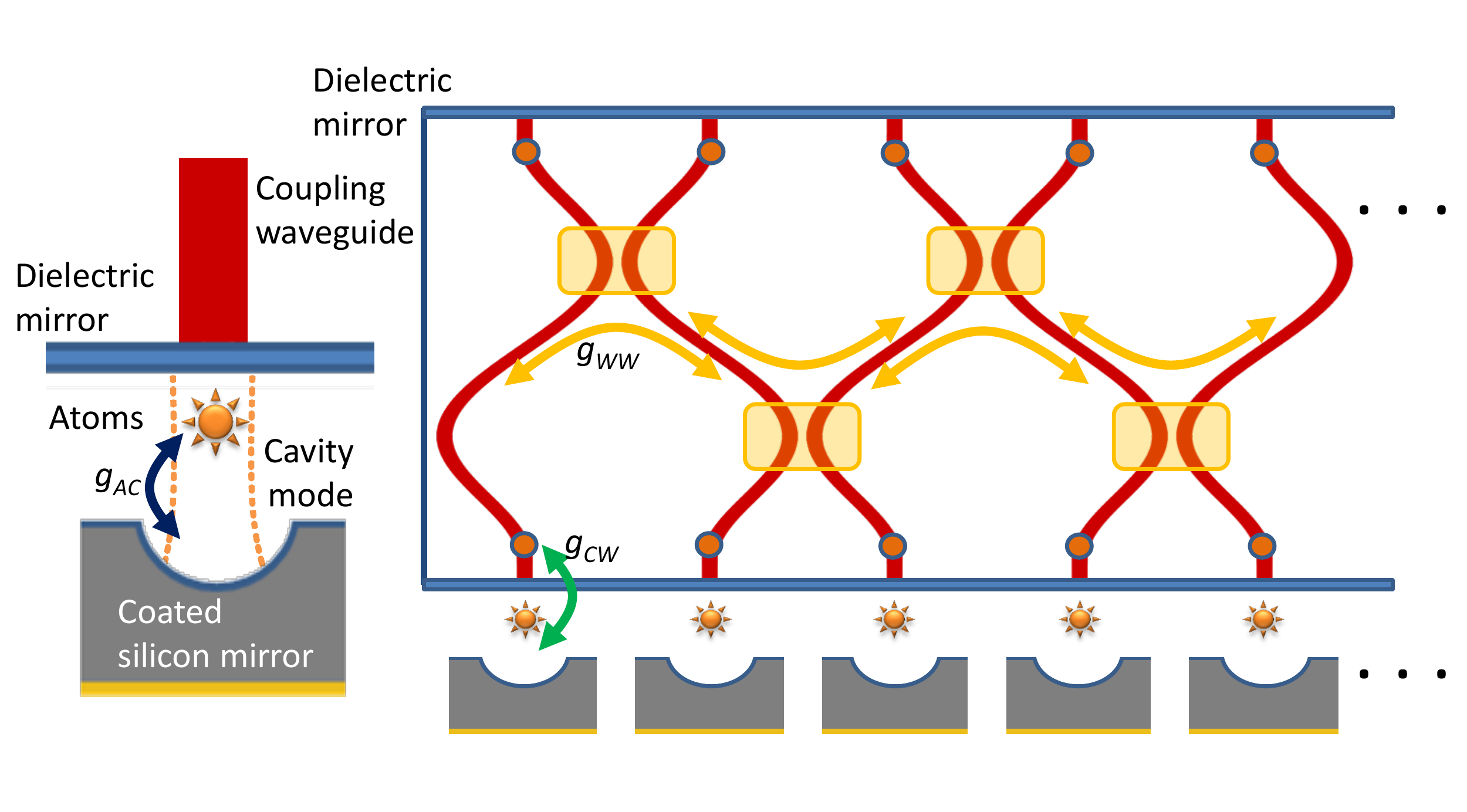}
\caption{Outline of the waveguide chip. W and C refer to the waveguide
cavities and the microcavities respectively.  Circles indicate the positions of active heating elements. These tune the resonant frequency of each waveguide, and adjust the coupling between adjacent waveguides.  Rectangles indicate the coupling regions. See Table \ref{definitionsandsymbols} for a description of the
symbols used.}
\label{fig:chip_outline}
\end{figure}

We envisage a device with quantum emitters in several separate microcavities, coupled by waveguides on a coupling chip as illustrated in Fig. \ref{fig:chip_outline}. The microcavities
are based on the design described in \cite{TGD+07}. Hemispherical micro-mirrors, wet-etched into a silicon chip~\cite{THE+05,TGD+07}
and reflection-coated, form one side of a set of Fabry-Perot microcavities. These cavities are closed by the
plane reflection-coated ends of waveguides integrated on a photonic chip. The other end of each waveguide is also reflection-coated to form a second set of optical resonators. Each waveguide/microcavity pair, coupled together through the shared mirror, forms a composite cavity. Henceforth, we will sometimes simply call this a cavity, and we will speak of a microcavity or a waveguide cavity if we wish to distinguish the component parts.  The microcavities are open in the transverse direction, giving access to lasers that trap and manipulate atoms at the position of maximum interaction with cavity mode. This design benefits from the intrinsic scalability of microfabrication to achieve controlled nearest-neighbour coupling of many optical cavities.
\begin{table}[h]
        \begin{tabular}{|c| p{5cm} |c|}
        \multicolumn{3}{c}{\bf Definitions and Symbols}\\
        \hline
        Symbol & Definition & Typical values\\
        \hline
        $L_W$, $L_C$ & length of the cavities & $L_C=\unit[(10-100)]{\mu m}$\\
        & & $L_W>\unit[(10-20)]{mm}$ \\
         $r_{i}$ & amplitude reflection & \\
        & coefficients of the mirrors & \\
        $R_{i} = |r_{i}|^{2}$ & Intensity reflection & $99.95\% > R_C= 99.9\%> 99.0\%$\\
        & coefficients of the mirrors& $99.95\% > R_W=99.0\%$ \\
        & & $R_{CW} \approx 98\% $\\
        \hline
        $FSR_i=\pi c/L_i$ & free spectral range & $FSR_C\approx \unit[2\pi \times (0.2-2)]{THz}$\\
        & & $FSR_W\approx \unit[2\pi\times (1-2)]{GHz}$\\
        $F_C=\frac{\pi}{1-\sqrt{R_C R_{CW}}}$ & cavity finesse & \\
        $F_W=\frac{\pi}{1-\sqrt{R_W R_{CW}}}$ & waveguide finesse & \\
        $\kappa_i=FSR_i/2 F_{i}$ & resonance linewidth & \\
        $w_C$ & microcavity mode waist & $w_C=\unit[3-5]{\mu m}$\\
        $w_W$ & waveguide mode size & ideally matched to $w_C$\\
        \hline
        $g_{AC} =\sqrt{\frac{3c \lambda^2\gamma}{\pi^2 w_C^2 L_C}}$ & atom-cavity coupling & $2\pi \times \unit[(0.1-1)]{GHz}$ \\
        $\gamma$ & atom amplitude decay rate & Rb: $\gamma=\unit[2\pi \times 3]{MHz}$\\
        $\lambda$ & wavelength & Rb: $\lambda=\unit[780]{nm}$\\
        \hline
        $J_{WW}$& tunnelling rate between adjacent waveguide resonators & $0<J_{WW}<\unit[2\pi \times 2]{GHz}$\\
        \hline
        \end{tabular}
        \footnotetext[2]{Typical working value}
\label{definitionsandsymbols}
\caption{Definitions of relevant symbols. All frequencies are angular: e.g. $2\pi \times \unit[1]{GHz}$ means $6.28\times10^{9}$s$^{-1}$.}
\end{table}

In the following discussion, we assume a wavelength of $\lambda = 780\,$nm, that being the D2 line of rubidium-87 atoms whose amplitude decay rate is $\gamma = 2\pi \times 3\,$MHz. In a microcavity of length $L_{C}$, the maximum coupling rate $g_{AC}$ between a two-level atom and the standing-wave field of one photon is
\begin{equation} \label{eq:cav-atom-coupl}
g_{AC} = \sqrt{\frac{3 c \lambda^{2} \gamma}{\pi^{2} w_{0}^{2} L_{C}}},
\end{equation}
where $w_{0}$ is the $1/e^{2}$ radius of the Gaussian intensity profile at the waist and $c$ is the speed of light. microcavities of the type envisaged can have
mode waists down to $2\mu$m, while the length of the microcavity $L_C$
is in the range of $10 - 100\,\mu$m. Therefore the coupling rate is of order $2 \pi \times 0.1 - 1$ GHz, and hence $g_{AC} \gg \gamma$. The reflectivity of each mirror is given
by $R_{i} = 1 - (T_{i} + A_{i})$. Assuming that the power transmission and absorption coefficients
fulfil $T_{i}, A_{i} \ll 1$, the cavity field amplitude decays at a rate
\begin{equation}\label{eq:kappa}
        \kappa_{C} = \frac{c \xi_{C}}{2 L_{C}} \quad \mathrm{with} \quad
        \xi_{C} \approx \frac{T_{C} + A_{C} + T_{CW} + A_{CW}}{2},
\end{equation}
 The subscript $C$ denotes the concave
microcavity mirror, while the subscript $CW$ denotes the plane coupling mirror between
the cavity and the waveguide. As we intend to fabricate the concave mirror
by isotropic etching of silicon, we expect the losses due to surface
roughness to be of order $A_{C} \approx 10^{-4}$ without additional polishing. The coupling
mirror, on the other hand, can have losses on the order of $A_{CW} \approx
5 \times 10^{-6}$, assuming the waveguide facet is super-polished and a
dielectric mirror formed by ion-assisted deposition is used. The decay
rate of the microcavity could then be made as small as $\kappa_{C}
\approx 2\pi \times 0.01$GHz for a $100\,\mu$m-long cavity. This length will give enough space to permit external optical access to the atoms. However, in order to couple effectively to the waveguide resonator, it is desirable to
increase the transmission of the coupling mirror so that $T_{CW} \gg T_{C}
+ A_{C}$. Therefore in practice, $\kappa_{C} \geq 2\pi \times 0.1$GHz.
This does not necessarily imply higher loss for the composite cavity since photons that go through this mirror enter the waveguide and are not necessarily lost.

Similarly, the field in the waveguide resonator decays at a rate
\begin{equation}
\kappa_{W} = \frac{c \xi_{W}}{2 L_{W}} \: \mathrm{with} \:
\xi_{W} \approx \frac{T_{W} + A_{W} + T_{CW} + A_{CW} + K_{W}}{2},
\end{equation}
where  $L_{W}$ is the optical length of the waveguide cavity (refractive index times physical length) and $K_{W}$ accounts for the waveguide propagation loss over one round-trip. While there is little use of integrated waveguides at $780\,$nm, we find in the literature \cite{MAT+02} that propagation losses can be less than $0.02\,$dB/cm for wavelengths around $800\,$nm in polymer waveguides. This platform offers all the necessary technological components for our device. We estimate that to achieve the necessary coupling lengths and separations between the waveguides without incurring additional bend losses, the waveguide resonator will have a length of $L_{W} > 1\,$cm. We will therefore conservatively consider $L_{W} = 2\,$cm, which gives a fractional round-trip loss of $1.8\,$\% and a corresponding decay rate of $\kappa_{W} > 2\pi \times 10 $MHz.

The coupling from the microcavity into the waveguide resonator depends on the transmission $T_{CW}$, but also on the spatial overlap of the waveguide and microcavity modes, which will need to be optimised experimentally. In principle, this can reach unity, and we estimate that at least 90\% should be achievable in practice. The photon tunnelling rate between microcavity and waveguide can easily exceed the free spectral range of the waveguide cavity. In this case, it is appropriate to consider the eigenmodes of the composite cavity, rather than viewing the microcavity and waveguide cavity as individual devices. We calculate the spectrum of one such coupled cavity in section \ref{sec:spectrum}.

\begin{figure}[t]
	\includegraphics[width=0.8\textwidth]{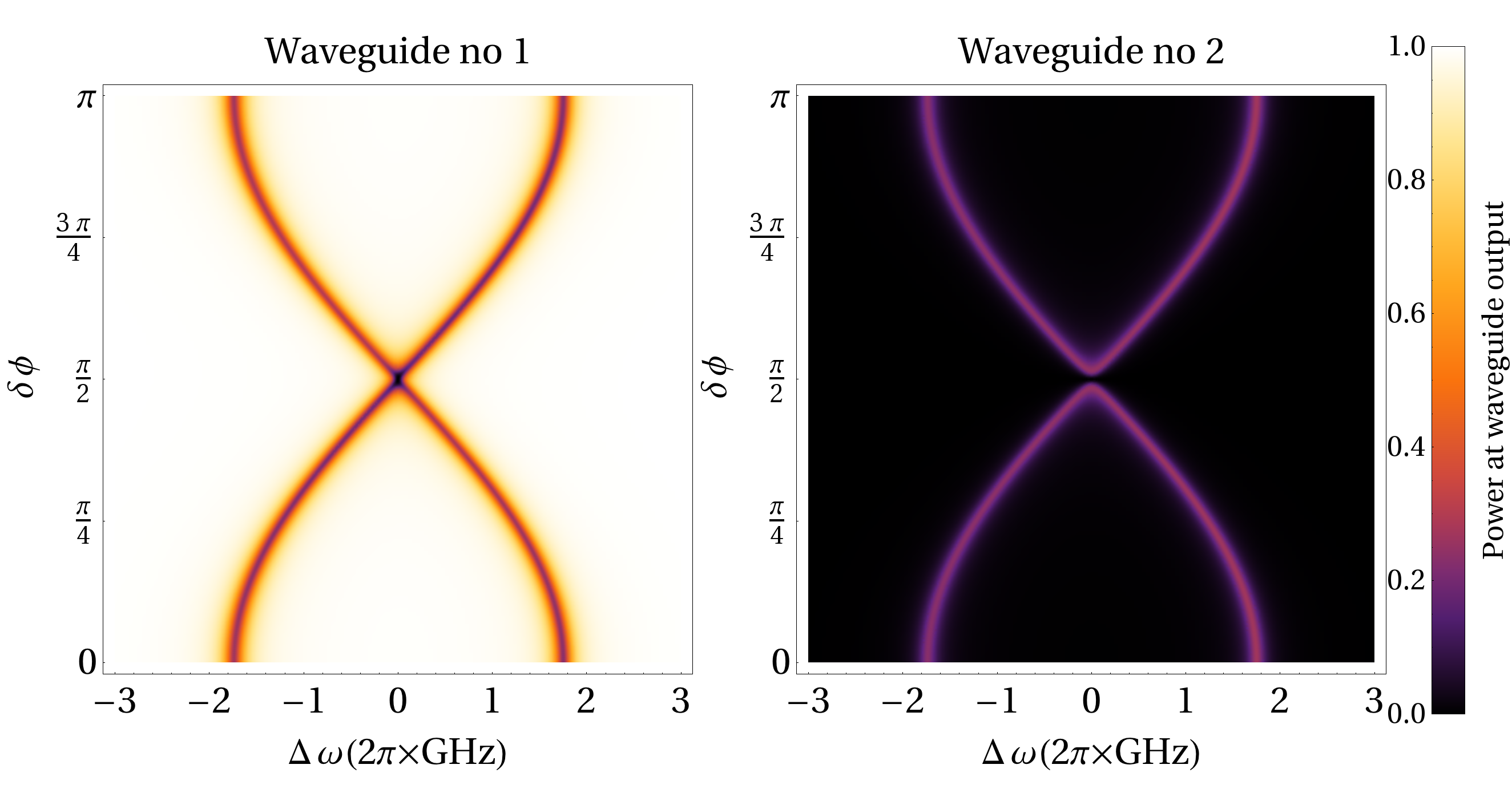}
	\caption{Calculated reflection spectrum of two composite cavities having evanescent-wave coupling between their waveguides. A pair of thermo-optics phase shifters is adjusted to give a phase shift $\delta \phi$ between the standing waves in the two waveguides. The system is pumped on waveguide number 1. At $\delta \phi = \pi/2$, the overlap integral of the standing waves is zero, so the coupling is switched off. Consequently, there is no normal mode splitting and no power is seen in waveguide number 2. When the phase shift changes to $\delta \phi = 0$ or $\pi$, the coupling is maximised and therefore so is the mode splitting. This calculation is based on an expanded version of the transfer matrix method described in \cite{Yariv07}.}
	\label{fig:cwGww}
\end{figure}

Each waveguide is coupled to its two nearest neighbours by short regions of evanescent field overlap to produce a tunnelling rate $J_{WW}$. The maximum value $J^{(max)}_{WW}$, determined by the maximum field overlap and the length of the coupling region, can be as high as $10^{10}$s$^{-1}$. This overlap can be tuned using thermal phase shifters, one near each end, to move the standing wave without changing the cavity length. In this way, $J_{WW}$ can be tuned in principle over the range $0 - J^{(max)}_{WW}$. In practice, a small amount of crosstalk will be unavoidable because of scattering, tuning noise and linewidth effects, but we estimate that $J^{(min)}_{WW} < 10^{-3} \times J^{(max)}_{WW}$. The maximum coupling strength must therefore be chosen judiciously to ensure access to the desired range of coupling.
The thermo-optic phase shifters also allow the optical length of each waveguide cavity to be tuned at a rate of up to a few wavelengths per ms. Figure \ref{fig:cwGww} illustrates how two coupled cavities exhibit a normal mode splitting and shows how the evanescent-wave coupling can be adjusted by shifting the phase of one standing wave relative to the other. In particular, the coupling is switched off for a relative phase of $\pi/2$.

\section{The composite cavity spectrum \label{sec:spectrum}}

As our starting point, we consider the spectrum of a single composite cavity. Several such cavities are to be concatenated in our setup to form the coupled-cavity array.

Suppose that we couple light into one of the cavities, through the top waveguide mirror as drawn in Fig.~\ref{fig:chip_outline}. Without the curved microcavity mirror at the bottom, the field reflected by the waveguide cavity is given by
\begin{equation}\label{eq:ErW}
E_{r,W} = \frac{r_{W}-r_{CW} e^{2i\phi_W}}{1-r_{W}r_{CW} e^{2i\phi_W}} \, E_{in} \, ,
\end{equation}
where $E_{in}$ is the incident field, $r_W$ and $r_{CW}$ are the amplitude reflection coefficients of the two mirrors, and $2\phi_W=4\pi L_W/\lambda$ is the round-trip propagation phase. Here we have assumed lossless mirrors for simplicity. Resonance occurs when $\phi_W=m\pi$ with $m$ being an integer. Similarly, a travelling field $E_{in,C}$ that is incident on the microcavity produces a reflected field in the waveguide given by
\begin{equation}
\frac{E_{r,C}}{E_{in,C}} = \frac{r_{CW}-r_C e^{2i\phi_C}}{1-r_{CW}r_{C} e^{2i\phi_C}} \equiv \tilde{r}_{C} e^{i\theta}\,,
\end{equation}
where $2\phi_C=4\pi L_C/\lambda$. This ratio defines the reflection coefficient $\tilde{r}_{C}$ and phase shift $\theta$ at the bottom end of the waveguide cavity when the microcavity mirror is in place:
\begin{eqnarray}\label{eq:theta}
\tilde{r}_{C}^2 & = & 
\frac{r_{CW}^2+r_C^2-2r_{CW}r_C\cos(2\phi_C)}
    {1+r_{CW}^2r_C^2-2r_{CW}r_C\cos(2\phi_C)}\,,\,\,\, \mbox{and}\nonumber\\
\theta & = & 
\arctan\left(\frac{(r_{CW}^2-1)r_C\sin(2\phi_C)}{r_{CW}(r_{C}^2+1)-r_{C}(r_{CW}^2+1)\cos(2\phi_C)}\right).
\end{eqnarray}
On replacing $r_{CW}$ in Eq.(\ref{eq:ErW}) by $\tilde{r}_{C} e^{i\theta}$, we obtain the reflected field from a composite cavity:
\begin{equation}
E_{r,CW} = \frac{r_{W}-\tilde{r}_{C}e^{i(\theta+2\phi_W)}}{1-r_{W}\tilde{r}_{C}e^{i(\theta+2\phi_W)}} \, E_{in}\,,
\end{equation}
which is in resonance when
\begin{equation}\label{eq:resCond}
2 \phi_W + \theta = 2 \pi n\,,
\end{equation}
for any integer $n$.
This set of equations describes the system fully. It cannot generally be solved analytically, but we can determine when the reflected intensity becomes zero. This is of interest because the spectrum is generally sensitive  to the presence of atoms under this condition. Zero reflected field is achieved when the microcavity can be tuned to make $r_{w}^{2} = \tilde{r}_{C}^{2}$. Such tuning is possible when the following inequality is satisfied:
\begin{equation}
|E_{r,C}^{min}|^2 = \frac{(r_{C}-r_{CW})^2}{(1-r_{C}r_{CW})^2}\leq r_W^2 \leq |E_{r,C}^{max}|^2=\frac{r_{C}^2+r_{CW}^2}{1+(r_{C}r_{CW})^2}\,,
\end{equation}
and occurs when the cavity round-trip phase satisfies
\begin{equation}\label{eq:OptDetuning}
2\phi_C(\mbox{opt}) =
\arccos{\left(
\frac{r_W^2(1+(r_{C}r_{CW})^2)-(r_{C}^2+r_{CW}^2)}
{2r_{CW}r_{C}(r_W^2-1)}
\right)}\,.
\end{equation}
This determines the phase shift $\theta$, and the resonance of the whole cavity is then ensured by adjusting the length of the waveguide to satisfy Eq.~(\ref{eq:resCond}).

\begin{figure}[t]
\centering
\includegraphics[width=0.8 \textwidth]{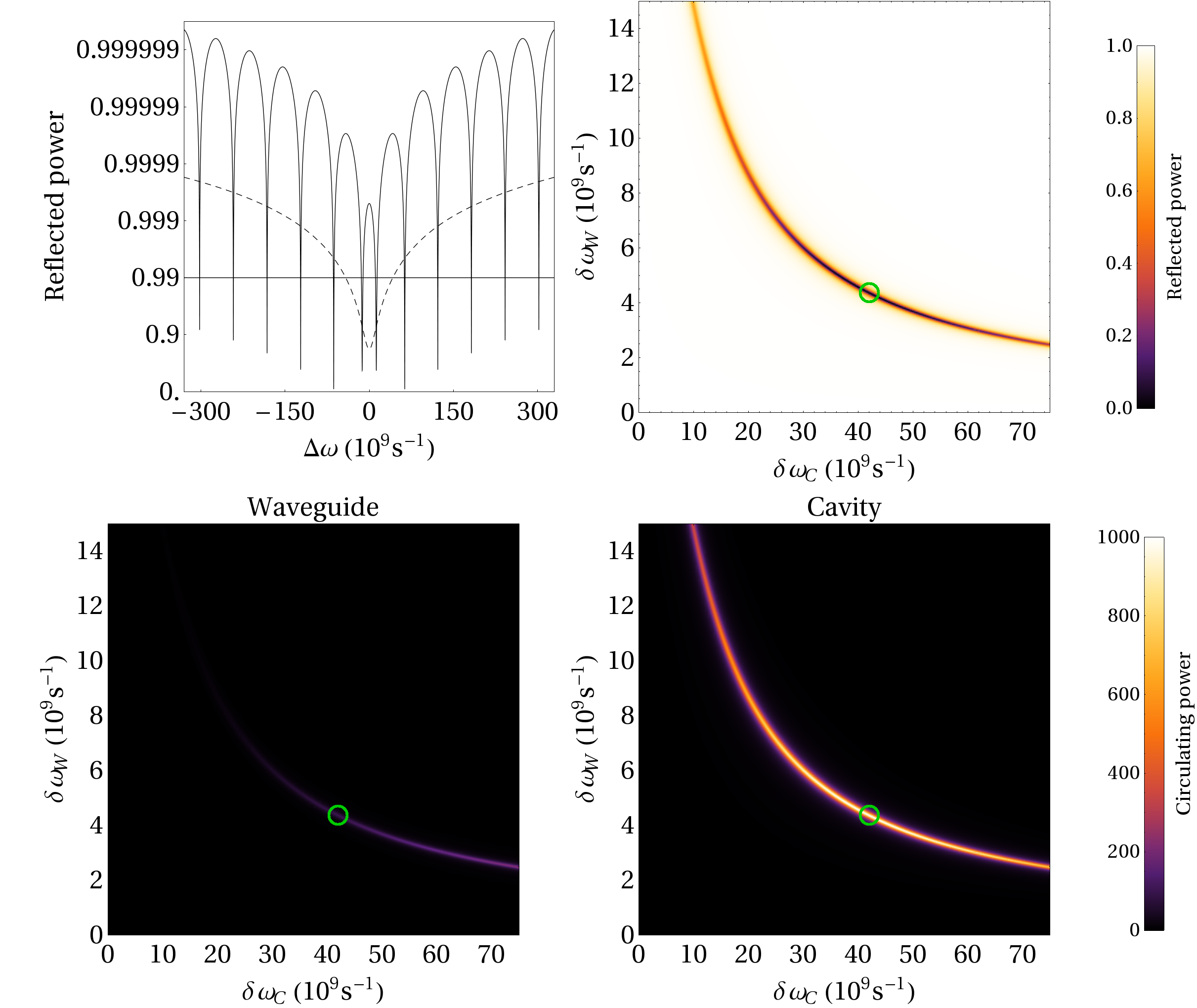} \caption{Behaviour of composite cavity with cavity lengths $L_C=156\,\mu$m, $L_W=15.6\,$cm, and power reflectivities $R_C=0.999$, $R_{CW}=0.98$, and $R_C=0.99$. Top left: Calculated reflection spectrum of the composite cavity versus detuning of the laser. In the absence of coupling, each separate cavity  has a resonance at $\Delta \omega=0$. Here they are coupled to produce normal modes split by $2.5\times10^{10}$s$^{-1}$. Dashed curve: $\tilde{r}_{C}^2$. Top right: reflected intensity near one of the central resonances. The laser frequency is fixed at $\Delta \omega=0$, while the (uncoupled) resonance frequencies of the two constituent cavities are varied. The green circle indicates the point of full contrast where the reflected intensity is zero, as calculated from Eq.
\ref{eq:OptDetuning}. Bottom: Circulating field strength inside the waveguide resonator (left) and the microcavity (right), as a function of their detuning from the laser. Green circles as as above.
}
\label{fig:cwSpectrum}
\end{figure}

The semi-logarithmic graph on the top left of Fig.~\ref{fig:cwSpectrum} shows how the reflected intensity varies with a change $\Delta\omega$ in the angular frequency of the laser for a cavity with $L_W=15.6\,$mm and $L_C=0.156\mu$m, and with mirror reflectivities of $R_{C}=0.999$, $R_{CW}=0.98$, $R_{W}=0.99$. Without any coupling, both constituent cavities are resonant at $\Delta\omega=0$. In the reflection spectrum we see a series of dips, symmetrically disposed around $\Delta\omega=0$. The two central dips represent normal modes of the composite cavity. These would be degenerate in the absence of coupling but are split apart here. Further away from this doublet, the resonances are not so strongly coupled to the microcavity and their spacing approaches the free spectral range of the bare waveguide cavity. The horizontal line at $0.99$ shows the reflection coefficient $r_{W}^2$ of the input mirror, while the dashed line indicates the value of $\tilde{r}_{C}^2$. These are most nearly equal at the second reflection dip on each side, making those dips the strongest ones. The width of each resonance can be understood in a simple way because the waveguide resonator is very much longer than the microcavity. Consequently $\tilde{r}_{C}$ and $\theta$ are essentially constant over any given resonance line of the coupled system. By analogy with Eq.~(\ref{eq:kappa}) this gives the cavity damping rate as
\begin{equation}
\tilde{\kappa} \approx \frac{c}{2 L_{W}}\xi \:\quad \mathrm{with} \quad \:
\xi \approx 1 - r_{W} \tilde{r}_C,
\label{eq:kappaeff}
\end{equation}

The top right graph in Fig.~\ref{fig:cwSpectrum} explores how the reflected intensity near one of the central resonances depends on the detuning of the two constituent cavities. We see that the reflection goes through a minimum whose position, indicated by the small circle, is as expected from Eq.(\ref{eq:OptDetuning}). The two lower graphs show how the circulating power inside the waveguide resonator (left) and the microcavity (right) varies with the same detuning of the constituent cavities. In particular, we see that the condition of minimum reflected power (again indicated by a circle) corresponds closely with having maximum power inside the microcavity, as required for good atom-cavity coupling. At this point, the energy density in the microcavity is ten times that in the waveguide (though the total energy is ten times less because of the hundredfold disparity in lengths).

\section{Approximation by a Jaynes-Cummings lattice Hamiltonian \label{sec:JC}}

In this sections we describe a series of approximations that allow us to obtain a description of our system in terms of atom photon interactions of Jaynes-Cummings form in each cavity together with tunnelling of photons between adjacent cavities.

\subsection{Composite cavity Hamiltonian}
The Hamiltonian for the field in one composite cavity is
\begin{equation}
H_{\mathrm{cav}} = \sum_{\alpha} \omega_{\alpha} a_{j,\alpha}^{\dagger} a_{j,\alpha},
\end{equation}
where the index $j$ labels the particular cavity and index $\alpha$ labels its eigenmodes.
The frequencies $\omega_{\alpha}$ are those of the eigenmodes discussed in Sec.~\ref{sec:spectrum}.

\subsection{Photon tunneling between adjacent waveguides}

The rate $J_{WW}$ for photons to tunnel between two adjacent waveguides can be tuned over a wide range. Let us choose to make it small compared with the waveguide free spectral range, which also ensures that $|J_{WW}| \ll |\omega_{\alpha} - \omega_{\alpha'}|$. In this regime, we can write the coupling as a tunnelling term between resonant modes of the two composite cavities,
\begin{equation*} 
H_{\mathrm{WW}} = - \sum_{\alpha, \beta} J_{\alpha, \beta} \left(a_{j,\alpha}^{\dagger} a_{j+1,\beta} + \mathrm{H.c.} \right) \approx - \sum_{\alpha} J_{\alpha, \alpha} \left(a_{j,\alpha}^{\dagger} a_{j+1,\alpha} + \mathrm{H.c.} \right)
\end{equation*}
When the cavity network only contains photons that are near-resonant with the normal modes $\alpha_{0}$,
and when the Rabi frequencies of atom-photon coupling are also  $\ll |\omega_{\alpha} - \omega_{\alpha'}|$, only these modes are populated and $H_{WW}$ can be simplified further to read,
\begin{equation}\label{eq:int_wg-wg}
H_{\mathrm{WW}} \approx - J_{\alpha_{0}, \alpha_{0}} \left(a_{j,\alpha_{0}}^{\dagger} a_{j+1,\alpha_{0}} + \mathrm{H.c.} \right).
\end{equation}
The tunnelling rate $J_{\alpha_{0}, \alpha_{0}}$ is related to $J_{WW}$. Yet, since $J_{WW}$ is harder to determine experimentally than $J_{\alpha_{0}, \alpha_{0}}$, we do not specify the conversion here.

\subsection{Atom-photon coupling}

The circulating fields in the microcavity and waveguide, $E_{C}$ and $E_{W}$ respectively, are given by,
\begin{eqnarray}
E_W & = & E_{\mathrm{in}} \frac{t_W}{1-r_W \tilde{r}_C e^{i \left(2 \phi_W+\theta \right)}} \label{eq:EW}\\
E_C & = & E_{\mathrm{in}} \frac{t_{CW} t_W e^{i \phi_W}}{1-r_C  r_{CW} e^{2 i \phi_C}+r_C r_W e^{2 i \left(\phi_C+\phi_W\right)}-r_{CW} r_W e^{2 i \phi_W}}\label{eq:EC}
\end{eqnarray}
where $E_{\mathrm{in}}$ is the input field from the top end of the waveguide, see figure \ref{fig:chip_outline}. Since one photon in the composite cavity has a total energy content of $ \hbar \omega$, the absolute values of the circulating field amplitudes per photon in the microcavity and waveguide are
\begin{eqnarray}
\mathcal{E}_C & = & \sqrt{\frac{\hbar \omega}{2 \varepsilon_{0} \pi w_{0}^{2}}}
\sqrt{\frac{\left|E_{C}\right|^{2}}{L_{C} \left|E_{C}\right|^{2} + L_{W} \left|E_{W}\right|^{2}}} \\
\mathcal{E}_W & = & \sqrt{\frac{\hbar \omega}{2 \varepsilon_{0} \pi w_{0}^{2}}}
\sqrt{\frac{\left|E_{W}\right|^{2}}{L_{C} \left|E_{C}\right|^{2} + L_{W} \left|E_{W}\right|^{2}}} .
\end{eqnarray}
Hence the atom-photon coupling at an antinode of the microcavity is
\begin{equation}\label{eq:gac-rescale}
g = g_{AC} \, \sqrt{\frac{\left|E_{C}\right|^{2}}{\left|E_{C}\right|^{2} + \frac{L_{W}}{L_{C}} \left|E_{W}\right|^{2}}}
\end{equation}
where $g_{AC}$ is given in eq. (\ref{eq:cav-atom-coupl}) for a 2-level atom. This formula for a two-level atom applies because we are considering the closed cycling transition $5^{2}S_{1/2}(F=2,m_F=\pm2) \leftrightarrow 5^{2}P_{3/2}(F'=3,m'_F=\pm3)$ on the $D_{2}$ line of $^{87}$Rb.

\subsection{Jaynes-Cummings lattice}

With the above approximations, the dynamics of atoms coupled to the composite cavity normal modes $a_{j,\alpha_{0}}$, can be described by a Jaynes-Cummings lattice model. Dropping the index $\alpha_{0}$ on the field operators, the Hamiltonian for $N$ composite cavities reads,
\begin{eqnarray} \label{JCarray}
H_{\mathrm{JCarray}} & = & \omega_{A} \sum_{j=1}^{N} \sigma_{j}^{+} \sigma_{j}^{-} + \omega_{C} \sum_{j=1}^{N} a_{j}^{\dagger} a_{j} \nonumber \\
& + & g \sum_{j=1}^{N} \left(\sigma_{j}^{+} a_{j} + \sigma_{j}^{-} a_{j}^{\dagger} \right) \\
& - & J \sum_{j=1}^{N-1} \left(a_{j}^{\dagger} a_{j+1} + a_{j} a_{j+1}^{\dagger} \right) \nonumber .
\end{eqnarray}
Here, $\omega_{C} = \omega_{\alpha_{0}}$, $\sigma_{j}^{-} = |g_{j}\ket \bra e_{j}|$ is the transition operator
between the excited state $|e_{j}\ket$ and ground state $|g_{j}\ket$ of the
atom in cavity $j$ and $a_{j}$ is the annihilation operator for photons in mode $\alpha_0$
of that cavity. The losses for the system described by this
Hamiltonian arise through spontaneous emission from the excited
states of the atoms at rate $\gamma$ and photon loss from the normal
modes $a_{j}$ at rate $\tilde{\kappa}$ given by Eq. (\ref{eq:kappaeff}).

\section{Spectroscopy for a driven Jaynes-Cummings array \label{sec:JCspectroscopy}}

We now consider an array of composite cavities connected to each other by nearest-neighbour coupling and excited at the end of one waveguide by a laser tuned to the normal mode $a_1$. The resonant pumping is described by an additional term
$H_{D} = \frac{\eta}{2} a_{1}^{\dagger} + \frac{\eta^{*}}{2} a_{1}$
in the Hamiltonian whereas photon leakage and spontaneous emission losses are taken into account by Markovian damping terms.
The dynamics of the resulting driven dissipative system is then described by the master equation,
\begin{eqnarray}
\frac{d\rho}{dt} = -i \left[H,\rho\right] & + & \gamma \sum_{j} ( 2 \sigma_{j}^{-} \rho \sigma_{j}^{+} - \sigma_{j}^{+}\sigma_{j}^{-} \rho - \rho \sigma_{j}^{+}\sigma_{j}^{-}) \nn \\
& + & \tilde{\kappa} \sum_{j} ( 2 a_{j} \rho a_{j}^{\dagger} - a_{j}^{\dagger} a_{j} \rho - \rho a_{j}^{\dagger} a_{j}),
\end{eqnarray}
where $H = H_{D} + H_{\mathrm{JCarray}}$. For one cavity, a driving amplitude $\eta$ gives a steady state energy in the cavity of $\hbar\omega\bra a^{\dagger}a\ket=\hbar\omega|\eta|^2/(2 \tilde{\kappa}^2)$. Also, the ratio of energy in the cavity to input power $P$ is $2\left(|E_{C}|^{2}L_C+|E_{W}|^{2}L_W\right)/(|E_{\mathrm{in}}|^2c)$, where $E_{C}$ and $E_{W}$ are the circulating fields in the cavity.  Hence, the power $P$ of the driving laser is related to the driving amplitude $\eta$ by,
\begin{equation} \label{eq:input-power}
P = \hbar \omega_{L} \, \frac{|\eta|^2}{\tilde{\kappa}^{2}} \,
\frac{E_{\mathrm{in}}^2c}{2\left(|E_{C}|^{2}L_C+|E_{W}|^{2}L_W\right)}.
\end{equation}
Below, we calculate the spectrum and photon statistics that can be observed at the output ports of the coupled waveguides \cite{Hartmann10,Leib10,Knap10,Ferretti10}.
These properties correspond to the most straightforward experiments that might be made using such a cavity array. As we shall see, they nonetheless reveal interesting physics including significantly entangled photon output states.

\subsection{Composite cavity modes with strong coupling and high single atom cooperativity}

\begin{figure}[t]
\centering
\includegraphics[width=0.8\textwidth]{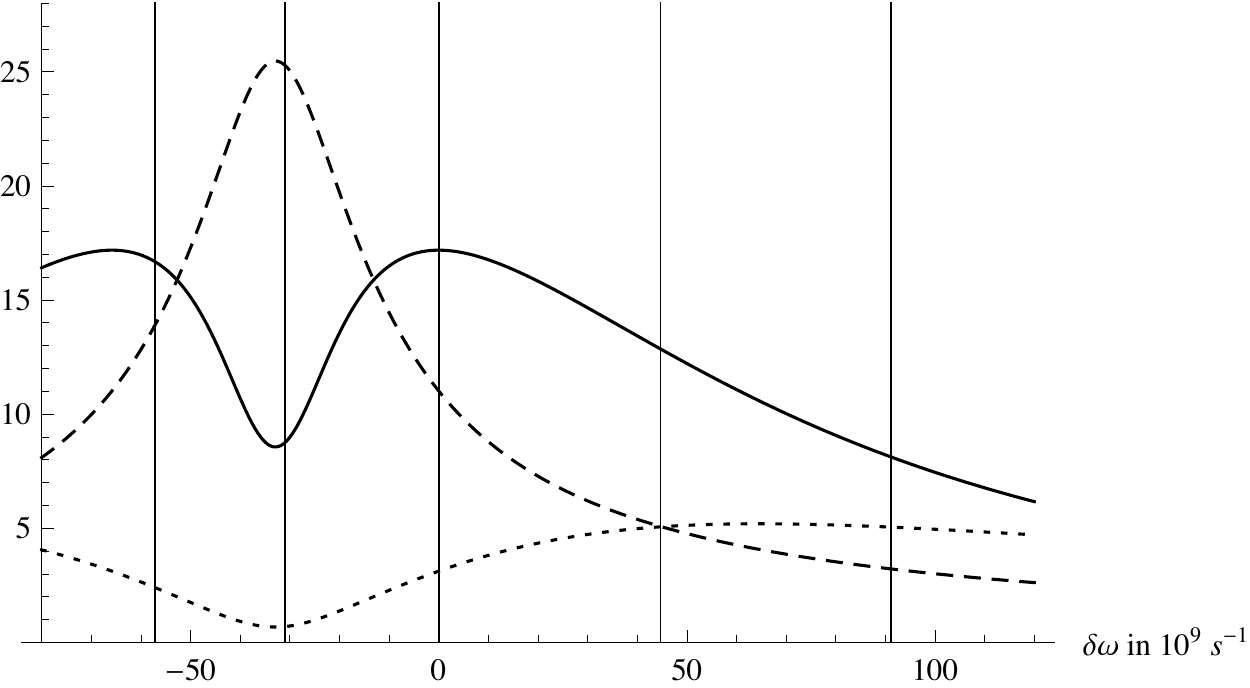}
\caption{Composite cavity properties versus laser detuning. The atomic transition frequency is at $\delta\omega=0$. The cavity parameters ($L_{C} = 156.05 \, \mu$m, $L_{W} = 20.000 \, $mm, $R_{C} = 99.9$\%, $R_{CW} = 98.0$\% and $R_{W} = 99.8$\%) are close to those of Fig.\,\ref{fig:cwSpectrum}. Vertical lines: frequencies of reflection minima due to composite cavity resonances. Solid line: single atom cooperativity $C_{1} = g^{2}/(2 \tilde{\kappa} \gamma)$. At its peak value $C_1 = 17.3$. Dashed line: coupling relative to atomic decay rate, $g/\gamma$. Dotted line: coupling relative to cavity decay rate,
$g/\tilde{\kappa}$.}
\label{fig:high-coop}
\end{figure}

Since we are interested in having large atom-photon coupling and small photon losses, both from the cavities and from spontaneous emission, we wish to work with a mode that has high single atom cooperativity, $C_1 = g^{2}/(2 \tilde{\kappa} \gamma)$ and is resonant with the atomic transition.  The required combinations of microcavity and waveguide lengths are found by numerical optimisation. In Fig.\,\ref{fig:high-coop} the cavity resonances are indicated by vertical lines. The resonance near $\delta\omega=0$, with the highest cooperativity (solid line) of $C_1 = 17.3$, coincides with the atomic transition frequency. The cavity parameters ($L_{C} = 156.05\,\mu$m, $L_{W} = 20.000\,$mm, $R_{C} = 99.9$\%, $R_{CW} = 98.0$\% and $R_{W} = 99.8$\%) are very close to the parameters used in Fig.\,\ref{fig:cwSpectrum}. This resonance corresponds to the strong reflection dip in Fig.\,\ref{fig:cwSpectrum} on the high frequency side of the central doublet. The atom-photon coupling $g$ and photon loss rate $\tilde{\kappa}$ are $g = 0.3304 \times g_{AC} = 2 \pi \times 33.04\,$MHz and $\tilde{\kappa} = 2 \pi \times 10.6\,$MHz.

\subsection{Steady-state of two-site Jaynes-Cummings array}\label{subsec:2JC steady}

\begin{figure}[b]
\includegraphics[width=\textwidth]{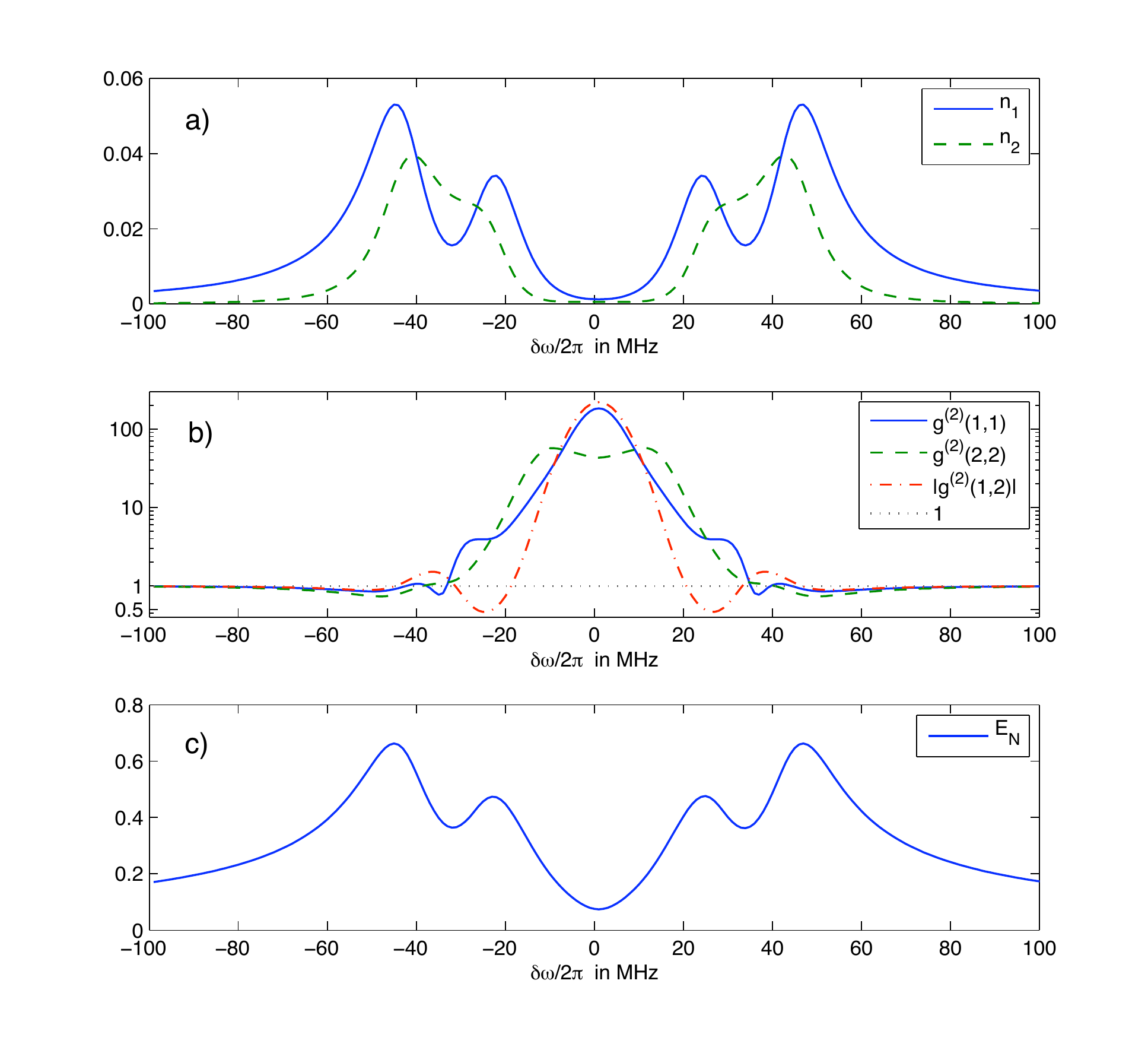}
\caption{Steady state of a driven array of two Jaynes-Cummings cavites for $g = 2 \pi \times 33.04\,$MHz, $\eta = 2 \pi \times 10\,$MHz, $J = 2 \pi \times 20\,$MHz, $\tilde{\kappa} = 2 \pi \times 10.6\,$MHz and $\gamma = 2 \pi \times 3 \,$MHz. {\bf a)} shows $n_{1}$ and $n_{2}$, {\bf b)} shows $g^{(2)}(1,1)$, $g^{(2)}(2,2)$ and $g^{(2)}(1,2)$ and {\bf c)} shows the entanglement between modes $a_{1}$ and $a_{2}$ as quantified by the logarithmic negativity $E_{N}$.}
\label{fig:steadyJC}
\end{figure}

Figure \ref{fig:steadyJC} shows the steady-state behaviour when two of these cavities are coupled together at a rate $J = 0.2 \times g_{AC} = 2 \pi \times 20\,$MHz. Each contains a rubidium atom and cavity number 1 is driven by a laser with $\eta = 0.1 \times g_{AC} = 2 \pi \times 10\,$MHz. The atoms are resonant with the normal modes $a_{j}$. Figure \ref{fig:steadyJC}a shows the number of photons in each cavity: $n_{1} = \bra a_{1}^{\dagger} a_{1} \ket$ and $n_{2} = \bra a_{2}^{\dagger} a_{2} \ket$. The four resonance lines in the spectrum correspond to the four singly excited eigenstates of the two-site Jaynes-Cummings Hamiltonian, i.e. Eq.(\ref{JCarray}) with  $N=2$. Here, the resonances at $\delta\omega \approx \pm \, 2\pi \times 44$ MHz correspond to states where the excitation is more likely to be found in one of the cavity modes whereas for the resonances at $\delta\omega \approx \pm \, 2\pi \times 24$ MHz the excitation is more likely to be in the atom than in the cavity field.

In figure \ref{fig:steadyJC}b we study the photon density correlations
$g^{(2)} (1,1) = \bra a_{1}^{\dagger} a_{1}^{\dagger} a_{1} a_{1} \ket / n_{1}^{2}$,
$g^{(2)} (1,2) = \bra a_{1}^{\dagger} a_{2}^{\dagger} a_{2} a_{1} \ket / n_{1} n_{2}$ and
$g^{(2)} (2,2) = \bra a_{2}^{\dagger} a_{2}^{\dagger} a_{2} a_{2} \ket / n_{2}^{2}$. On the two outer resonances ($\delta\omega \approx \pm \, 2\pi \times 44$ MHz) we see small dips below unity, indicating photon anti-bunching in both cavities, $g^{(2)} (1,1) <1$ and $g^{(2)} (2,2) <1$, as well as anti-correlations between photons in distinct cavities, $\left|g^{(2)} (1,2)\right| <1$ . This indicates that the coupling in both cavities is strong enough to generate an optical nonlinearity that converts coherent classical input light into a manifestly non-classical state of the cavity photons. For the resonances at $\delta\omega \approx \pm \, 2\pi \times 44$ MHz the state of the two cavity modes is approximately a superposition of one photon in cavity 1 with cavity 2 empty and one photon in cavity 2 with cavity 1 empty which gives rise to the anti-correlations shown by $g^{(2)} (1,1)$, $g^{(2)} (2,2)$ and $g^{(2)} (1,2)$. Since cavity 2 is not directly driven, it experiences a lower intensity of incoming photons than cavity 1 and hence exhibits slightly stronger anti-bunching for the same nonlinearity. For this state of the cavity modes, the photons in the two cavities are expected to become entangled. This is confirmed by figure \ref{fig:steadyJC}c which shows the entanglement between modes $a_{1}$ and $a_{2}$ as quantified by the logarithmic negativity $E_{N}$~\cite{PV06},
\begin{equation}
E_{N} = \log_{2} \left(\mathrm{Tr}\sqrt{\rho_{\mathrm{pt}}^{\dagger} \rho_{\mathrm{pt}}}\right),
\end{equation}
where $\rho_{\mathrm{pt}}$ is the partial transpose of the reduced density matrix of the two modes $a_{1}$ and $a_{2}$.

On the inner resonances ($\delta\omega \approx \pm \, 2\pi \times 24$ MHz), by contrast, only photons involving both cavities are anti-correlated whereas photons within each cavity bunch. This indicates that photons prefer to stick together in either of the cavities and the associated state contains superpositions of two or more photons in cavity 1 with cavity 2 empty and the reverse configuration. Hence photons in the output of one cavity tend to come in bunches whereas if one cavity emits a photon the other cavity is unlikely to emit at the same time. Consequently we also find entanglement between the photon modes for these resonances although less than for the resonances at $\delta\omega \approx \pm \, 2\pi \times 44$ MHz which feature higher photon densities.

Close to $\delta\omega = 0$, the photons in the cavities show pronounced bunching. This emerges due to the nonlinear spectrum of our device which here causes the laser drive to be detuned with respect to single photon transitions but resonant with multi photon transitions. In practice this bunching will however be hard to observe since the photon densities generated in these multi photon processes are vanishingly small as apparent from figure \ref{fig:steadyJC}a.

The findings shown in figures \ref{fig:steadyJC}b and c clearly demonstrate the non-classical nature of the light fields generated in our device.

\section{Effective Spin Hamiltonians \label{sec:effspin}}

\begin{figure}[b]
\centering
\includegraphics[width=8cm]{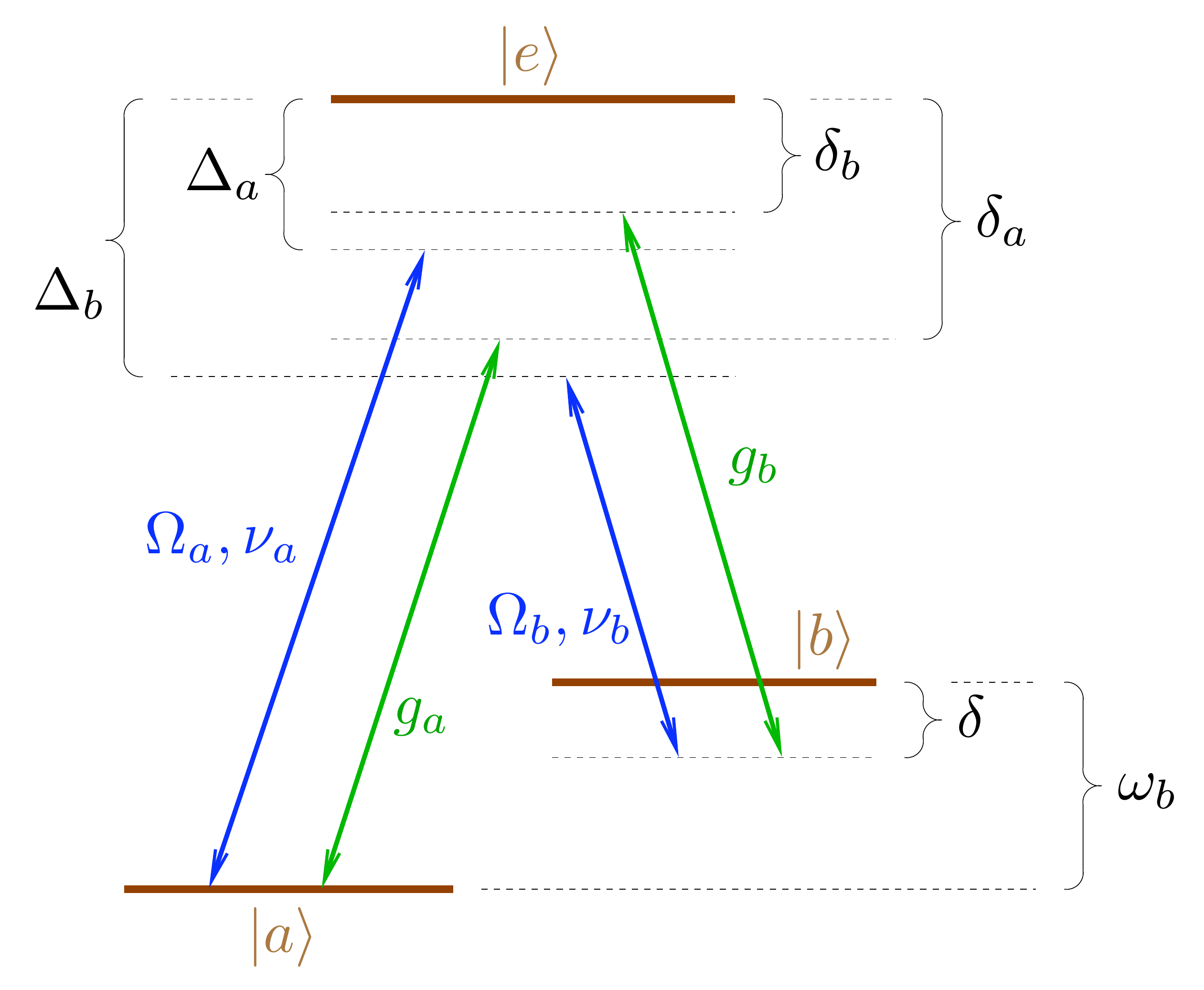}
\caption{Lambda level structure for the $D_{2}$ line of $^{87}$Rb atoms. The state $|a\ket$ represents  $5^{2}S_{1/2}(F=1,m_{F}=1)$, $|b\ket$ stands for $5^{2}S_{1/2}(F=2,m_{F}=1)$ and $|e\ket$ is $5^{2}P_{3/2}(F'=2,m_{F'}=2)$. The transitions $|a\ket \leftrightarrow |e\ket$  and $|b\ket \leftrightarrow |e\ket$ are driven both by the cavity field (green arrows) with coupling strengths $g_{a}$ and $g_{b}$, and by external coherent fields (blue arrows) with Rabi frequencies $\Omega_{a}$ and $\Omega_{b}$.
$\delta_{a} = \omega_{e} - \omega_{C}$ and $\delta_{b} = \omega_{e} - (\omega_{b} - \delta) - \omega_{C}$. State $|b\ket$ is stable as there is no electric dipole moment for the transition $|a\ket \leftrightarrow |b\ket$}.
\label{lambdalevel}
\end{figure}

In this section we describe a second experiment that could be performed using such a device. We show how effective spin-spin interactions, as proposed in \cite{HBP07a}, can be implemented. Here, each cavity interacts with one Rubidium atom, whose energy levels $a, b, e$ associated with the $D_{2}$ line form the lambda structure depicted in figure \ref{lambdalevel}. These levels are coupled both by the cavity photons (with couplings $g_{a}$ and $g_{b}$) and by external laser fields (with angular frequencies $\nu_a$, $\nu_b$ and Rabi frequencies $\Omega_{a}$, $\Omega_{b}$).
Importantly, the transition $|a\ket \leftrightarrow |b\ket$ ($5^{2}S_{1/2}(F=1,m_{F}=1) \leftrightarrow 5^{2}S_{1/2}(F=2,m_{F}=1)$) is dipole forbidden and level $|b\ket$ is thus metastable. Under conditions spelled out below, the dynamics can be constrained to the subspace formed by levels $|a_{j}\ket$ and $|b_{j}\ket$ of each atom and we can identify $|a_{j}\ket$ with spin down, $\mid\downarrow_{j}\ket \equiv |a_{j}\ket$, and $|b_{j}\ket$ with spin up, $\mid\uparrow_{j}\ket \equiv |b_{j}\ket$.

\subsection{Outline of the approach}

Following the arguments presented in section \ref{sec:spectrum}, we assume once again that the atom in the $j^{th}$ composite cavity only couples to one normal mode $a_{j}$ (again, we skip the index $\alpha_{0}$), for which we maximise the single atom cooperativity as in section \ref{sec:JCspectroscopy}. The photons tunnel between adjacent composite cavities at a rate $J$. The Hamiltonian for this system can thus be written as
\begin{eqnarray} \label{eq:ham-lambda-chain}
H & = & H_A + H_C + H_{AC} \quad \mathrm{where}\\
H_A & = & \sum_{j=1}^N \omega_e |e_j\ket \bra e_j| + \omega_{b} |b_j\ket \bra b_j|, \\
H_C & = & \omega_C \sum_{j=1}^N a_j^{\dagger} a_j - J \sum_{j=1}^N \left( a_j^{\dagger} a_{j+1} + a_j a_{j+1}^{\dagger} \right) \quad \mathrm{and}\\
H_{AC} & = & \sum_{j=1}^N \sum_{x = a,b} \left[ \left(\frac{\Omega_x}{2} e^{-i \nu_x t} + g_x a_j \right) |e_j\ket \bra x_j| + \mathrm{H.c.} \right].
\end{eqnarray}
Here, $H_A$ describes the atoms, $H_C$ the cavity field and $H_{AC}$ the interaction between atoms, lasers and cavity field. Also, $\omega_e$ is the $a-e$ transition frequency and $\omega_{b}$ the $a-b$ transition frequency.

The Hamiltonian $H_{C}$ can be decomposed into non-interacting collective photon modes, $H_{C} = \sum_{k} \omega_{k} a_{k}^{\dagger} a_{k}$, where $\omega_{k} = \omega_{C} - 2 J \cos k$ and $a_{k} = \sqrt{\frac{2}{N+1}} \sum_{j=1}^{N} \sin (k j) a_{j}$
with $k = \frac{\pi l}{N+1}$ and $l = 1, 2, \dots N$. It is helpful to move to an interaction picture,
\begin{equation}
H(t) = e^{i H_{0} t} (H-H_{0}) e^{-i H_{0} t}
\end{equation}
%
where H is given in equation (\ref{eq:ham-lambda-chain}) and $H_{0}$ reads,
\begin{equation}
H_0 = \sum_{j=1}^N \left( \omega_e |e_j\ket \bra e_j| + (\omega_{b} - \delta) |b_j\ket \bra b_j|\right) + \sum_{k} \omega_{k} a_k^{\dagger} a_k,
\end{equation}
with the value of $\delta$ to be chosen later.
In this picture, the Hamiltonian (\ref{eq:ham-lambda-chain}) becomes,
\begin{eqnarray} \label{eq:ham-lambda-chain-ip}
H(t) & = & \delta \sum_{j=1}^N |b_j\ket \bra b_j| + \\
& + & \sum_{j=1}^N \sum_{x = a,b} \sum_{k} \left[ \left(\frac{\Omega_x}{2} e^{i \Delta_x t} + g_{x,j,k} e^{i \delta_{k}^{x}x t} a_{k} \right) |e_j\ket \bra x_j| + \mathrm{H.c.} \right] \nonumber \, ,
\end{eqnarray}
where $\Delta_a = \omega_e - \nu_a$, $\Delta_b = \omega_e - (\omega_{b} - \delta) - \nu_{b}$, as illustrated in Fig.\,\ref{lambdalevel},
$\delta_{k}^{a} = \omega_e - \omega_{k}$ and $\delta_{k}^{b} = \omega_e - (\omega_{b} - \delta) - \omega_{k}$. The coupling constants $g_{a,j,k}$ and $g_{b,j,k}$ are related to the couplings $g_{a}$ respectively $g_{b}$ via $g_{x,j,k} = \sqrt{\frac{2}{N+1}} \sin (k j) g_{x}$ for $x = a, b$.

A judicious choice of rotating frame is $\delta=\omega_{b}-\frac{1}{2}(\nu_{a}-\nu_{b})$, which ensures that $\delta_{k}^{a} - \Delta_b = \delta_{k}^{b} - \Delta_a$ for all $k$ and allows a convenient separation of fast rotating terms from near-resonant terms in what follows. In this rotating frame, a second order adiabatic elimination of the excited levels $|e_{j}\ket$ and the photons $a_{j}$, see \ref{sec:adia_elim}, yields the effective spin-$\frac{1}{2}$ Hamiltonian,
\begin{equation} \label{eq:spinham}
H_{\mathrm{spin}} = B \sum_{j=1}^{N} \sigma_{j}^{z} + \sum_{j \neq l} \left(J_{j,l} \sigma_{j}^{+} \sigma_{l}^{+} + J_{j,l}^{\star} \sigma_{j}^{-} \sigma_{l}^{-} + K_{j,l} \sigma_{j}^{+} \sigma_{l}^{-} \right)
\end{equation}
where the effective transverse field $B$ reads,
\begin{eqnarray*}
B & = & \frac{\delta}{2} - \frac{|\Omega_{b}|^{2}}{8 \Delta_{b}^{2}} \left[ \Delta_{b} - \frac{|\Omega_{b}|^{2}}{2 \Delta_{b}} - \frac{|\Omega_{a}|^{2}}{4 (\Delta_{a} - \Delta_{b})} - \sum_{k} \left( \frac{|g_{b,j,k}|^{2}}{\delta_{k}^{b} - \Delta_{b}} + \frac{|g_{a,j,k}|^{2}}{\delta_{k}^{a} - \Delta_{b}} \right) \right] \\
& + & \frac{|\Omega_{a}|^{2}}{8 \Delta_{a}^{2}} \left[ \Delta_{a} - \frac{|\Omega_{a}|^{2}}{2 \Delta_{a}} - \frac{|\Omega_{b}|^{2}}{4 (\Delta_{b} - \Delta_{a})} - \sum_{k} \left( \frac{|g_{a,j,k}|^{2}}{\delta_{k}^{a} - \Delta_{a}} + \frac{|g_{b,j,k}|^{2}}{\delta_{k}^{b} - \Delta_{a}} \right) \right],
\end{eqnarray*}
and the coupling constants are
\begin{eqnarray*}
J_{j,l} & = & \frac{\Omega_{a} \Omega_{b}^{\star}}{4 \Delta_{a} \Delta_{b}} \sum_{k} \frac{g_{b,j,k}^{\star} g_{a,l,k}}{\delta_{k}^{b} - \Delta_{a}}, \\
K_{j,l} & = & \frac{|\Omega_{a}|^{2}}{4 \Delta_{a}^{2}} \sum_{k} \frac{g_{b,l,k}^{\star} g_{b,j,k}}{\delta_{k}^{b} - \Delta_{a}} +
\frac{|\Omega_{b}|^{2}}{4 \Delta_{b}^{2}} \sum_{k} \frac{g_{a,j,k}^{\star} g_{a,l,k}}{\delta_{k}^{a} - \Delta_{b}}.
\end{eqnarray*}
The Hamiltonian (\ref{eq:spinham}) is an appropriate description provided that $|\Delta_{x} - \delta_{y}| \ll FSR_{W}$ for $x,y = a,b$, $|\frac{\Omega_{x}}{2 \Delta_{x}}| \ll 1$ for $x = a,b$ and
$|\frac{\Omega_{x} g_{y,j,k}}{2 \Delta_{x}(\Delta_{x} - \delta_{k}^{y})}| \ll 1$ for $x,y = a,b$, see conditions (\ref{eq:adiaelimcond}). Furthermore, all the eigenmodes of the composite cavity should be sufficiently detuned from the $|e\ket\to|a\ket$ and  $|e\ket\to|b\ket$ transitions to avoid Purcell-enhanced atomic relaxation.

The sums over all photon modes, $\sum_{k}$, depend on the number of cavities in the array.
As a starting point for experiments, we focus here on the $N=2$ case. The explicit expressions for this case are given in \ref{sec:N2exprs}.

\subsection{Effective spin dynamics for two coupled cavities}

\begin{figure}[b!]
\centering
\includegraphics[width=\textwidth]{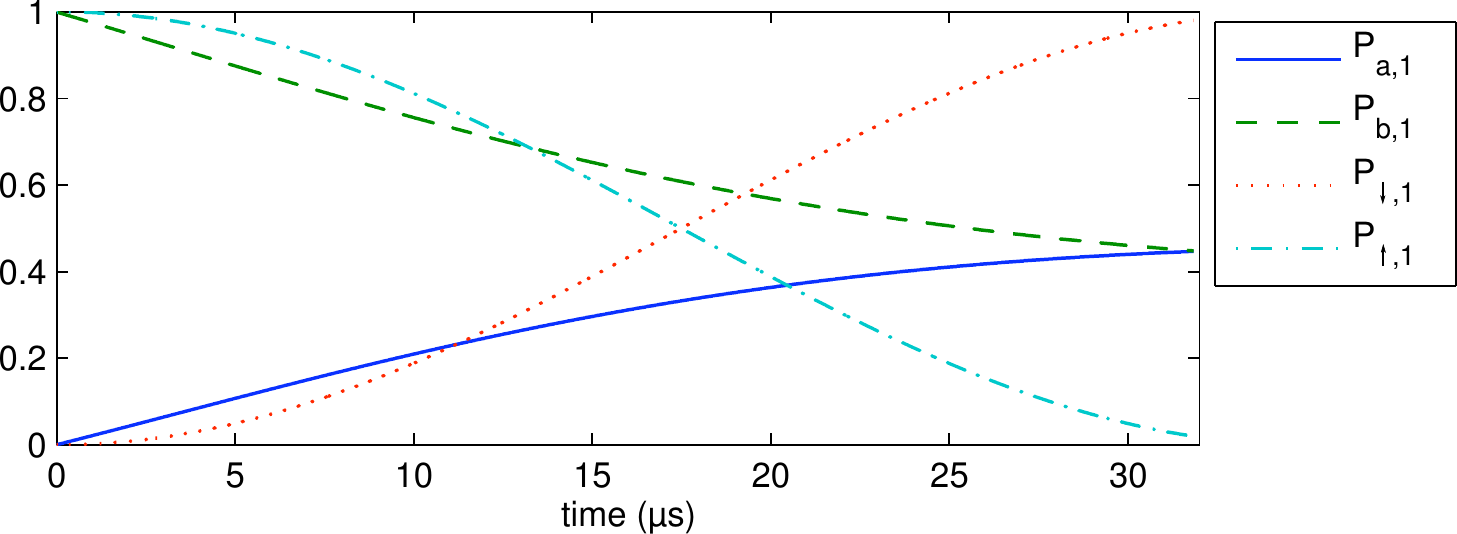}
\caption{Evolution of effective spin using the two-atom, two-cavity system described in Sec.~\ref{subsec:2JC steady}. The cavity details are given in the caption of Fig.~\ref{fig:high-coop}. Other parameters are
$J = 2 \pi \times 100.0 \, $MHz,
$g_{a} = 2 \pi \times 23.36 \, $MHz,
$g_{b} = 2 \pi \times 13.49 \, $MHz,
$\tilde{\kappa} = 2 \pi \times 10.59 \, $MHz,
$\Omega_{a} = 2 \pi \times 166.67 \, $MHz,
$\Omega_{b} = 2 \pi \times 394.16 \, $MHz,
$\Delta_{a} = 2 \pi \times 5.000 \, $GHz,
$\Delta_{b} = 2 \pi \times 11.825 \, $GHz,
$\delta_{a} = 2 \pi \times 11.740 \, $GHz,
$\delta_{b} = 2 \pi \times 4.915 \, $GHz and
$\delta = 2 \pi \times 10.0 \, $MHz.
These values lead to $B = 2 \pi \times 4.05 \, $MHz,
$J_{j,l} = 2 \pi \times 6.188 \, $kHz and
$K_{j,l} = 2 \pi \times 7.145 \, $kHz.
Dotted and dash-dotted lines show the probabilities for states $\mid\downarrow_{1}\ket$, $\mid\uparrow_{1}\ket$, according to Eq.(\ref{eq:spinham}). Dashed and solid lines show probabilities for states $|a_{1}\ket$ and $|b_{1}\ket$ according to the Master equation given in \ref{sec:full_lambda}. The damping has a significant effect.
}
\label{effspindamp2}
\end{figure}

As an example of dynamical evolution under this effective spin Hamiltonian, we consider once again the two coupled cavities described in Sec.~\ref{subsec:2JC steady}, each containing a $^{87}$Rb atom. The atom in cavity number 1 is prepared in state $|b_{1}\ket$ while the other atom is placed in state $|a_{2}\ket$ to form an initial effective spin state $\mid\uparrow_{1},\downarrow_{2}\ket$. The subsequent evolution of this state under the Hamiltonian of equation (\ref{eq:spinham}) is illustrated by the dotted and dash-dotted lines in Fig.\,\ref{effspindamp2}, which plot the probabilities of the spin-up and spin-down states in cavity 1:  $P_{\uparrow,1} = \mathrm{Tr}(\mid\uparrow_{1}\ket\bra\uparrow_{1}\mid \tilde{\rho})$ and
$P_{\downarrow,1} = \mathrm{Tr}(\mid\downarrow_{1}\ket\bra \downarrow_{1}\mid \tilde{\rho})$, $\tilde{\rho}$ being the state of the effective spin system. We see that the spin oscillates in this case with a period of approximately 70\,$\mu$s, determined by the coupling constant $K$ in equation (\ref{eq:spinham}). The symmetry of the problem ensures that $P_{\uparrow,2}=P_{\downarrow,1}$ and $P_{\downarrow,2}=P_{\uparrow,1}$. The values of $g_{a}$ and $g_{b}$ are derived from equation (\ref{eq:gac-rescale}) (via equations (\ref{eq:cav-atom-coupl}), (\ref{eq:EW}) and (\ref{eq:EC})), together with the appropriate weightings ($1/\sqrt{2}$ and $1/\sqrt{6}$) relative to the cycling transition. The values of these and the other relevant constants are listed in the caption of Fig.\,\ref{effspindamp2}.

\begin{figure}[b]
\centering
\includegraphics[width=\textwidth]{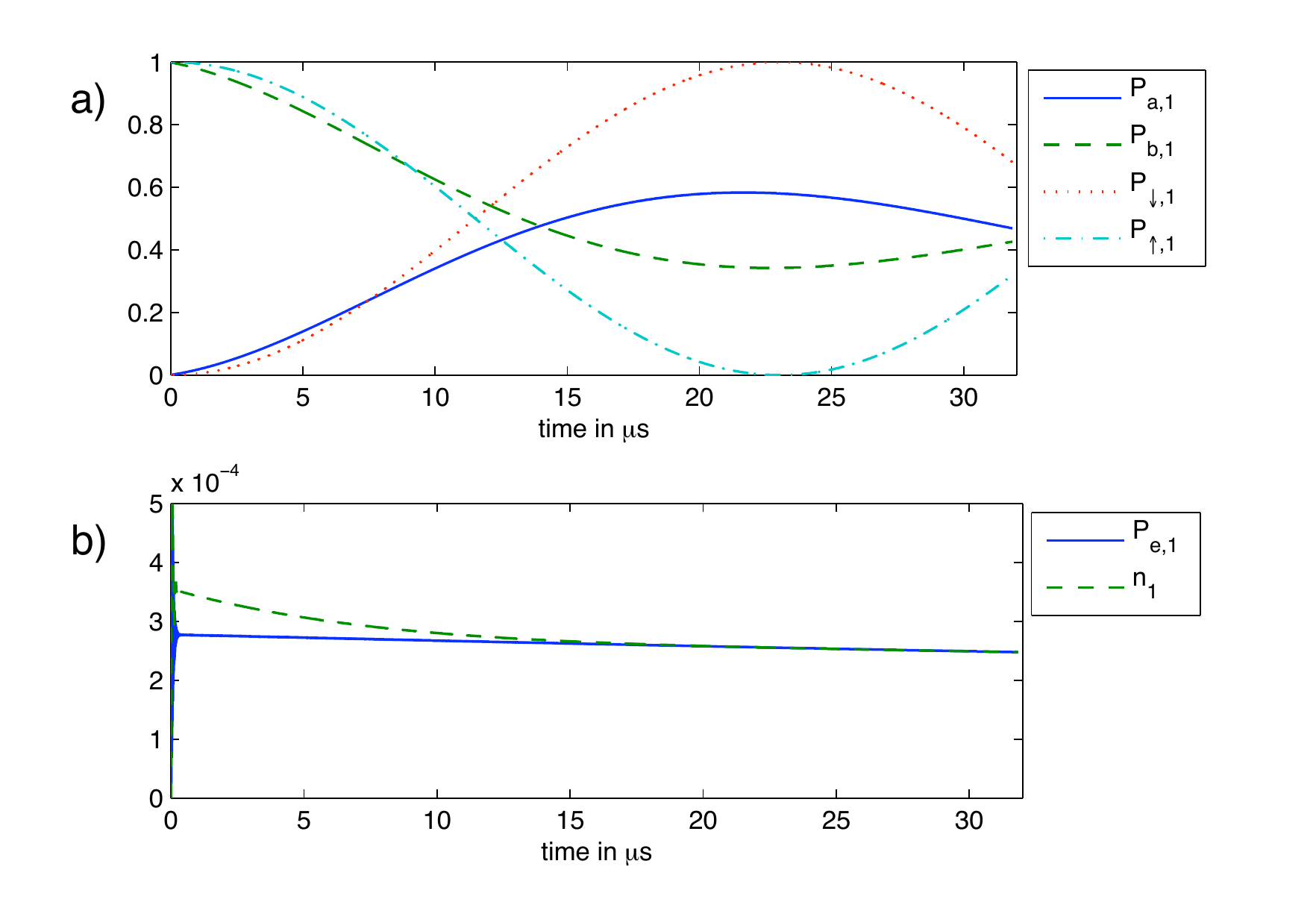}
\caption{Dynamics for effective spin system approach for
$R_{C} = 99.99$\%, $R_{CW} = 98.0$\%, $R_{W} = 99.9$\%, $L_{C} = 103.4 \mu$m, $L_{W} = 20.000$mm,
$J = 2 \pi \times 100.0 \, $MHz,
$g_{a} = 2 \pi \times 29.64 \, $MHz,
$g_{b} = 2 \pi \times 17.11 \, $MHz,
$\tilde{\kappa} = 2 \pi \times 3.10 \, $MHz,
$\Omega_{a} = 2 \pi \times 166.67 \, $MHz,
$\Omega_{b} = 2 \pi \times 394.16 \, $MHz,
$\Delta_{a} = 2 \pi \times 5.000 \, $GHz,
$\Delta_{b} = 2 \pi \times 11.825 \, $GHz,
$\delta_{a} = 2 \pi \times 11.741 \, $GHz,
$\delta_{b} = 2 \pi \times 4.916 \, $GHz and
$\delta = 2 \pi \times 10.0 \, $MHz.
These values lead to $B = 2 \pi \times 4.05 \, $MHz,
$J_{j,l} = 2 \pi \times 9.40 \, $kHz and
$K_{j,l} = 2 \pi \times 10.85 \, $kHz.
{\bf a)} Occupation probabilities $P_{x,1}$ for the states $\mid\downarrow_{1}\ket$, $\mid\uparrow_{1}\ket$, $|a_{1}\ket$ and $|b_{1}\ket$. {\bf b)} Occupation probability of the excited state $P_{e,1}$ and expectation value for the number of photons $n_{1}$. The initial state is $|b_{1},a_{2}\ket$ respectively $\mid\uparrow_{1},\downarrow_{2}\ket$.
}
\label{effspindamp3}
\end{figure}

In deriving equation (\ref{eq:spinham}) we adiabatically eliminated the excited state $|e\ket$ and therefore ignored the spontaneous emission from the atoms. In reality, this emission dephases the effective spin and causes loss of probability from the three-level system $\{a,b,e\}$. In addition, equation (\ref{eq:spinham}) ignores the loss of photons from the cavity. If a real experiment is to simulate the dynamics of a spin chain, as given by the Hamiltonian in equation (\ref{eq:spinham}), these rates of dissipation must be small enough. To test whether this is the case, we have calculated the full dynamics of the relevant atomic levels, including spontaneous emission and photon leakage using the  master equation given in \ref{sec:full_lambda}. The solid and dashed lines in Fig.\,\ref{effspindamp2} show the probabilities for the atom in cavity 1 to be in states $a$ and $b$:  $P_{a_{1}} = \mathrm{Tr}(|a_{1}\ket\bra a_{1}|\tilde{\rho})$ and $P_{b_{1}} = \mathrm{Tr}(|b_{1}\ket\bra b_{1}|\tilde{\rho})$. These should correspond to the spin-up and -down states of the equivalent spin model, but we see that they do not because the coherence is damped, leaving an incoherent mixture of states $|a\ket$ and $|b\ket$.

The situation can be rectified, as illustrated in Figure\,\ref{effspindamp3}(a). Here, the reflectivity of the microcavity mirror has been increased from $99.9\%$ to $99.99\%$ (together with a slight shortening of the microcavity and a minor improvement in the waveguide mirror). This improvement in the microcavity mirror allows the real system to provide a reasonable approximation to the ideal spin evolution, albeit with some residual damping. Figure\,\ref{effspindamp3}(b) shows the shows the excited state population $P_{e,1}$ on site number 1, which determines the lifetime $(P_{e}\times \frac{1}{3}\times 2\gamma)^{-1}\simeq260\,\mu$s for atoms to be lost from the lambda system by spontaneous decay to the state $5^{2}S(F=2,m_F=2)$. Also shown is the small population $n_{1}$ of cavity photons. Significant further improvement in the coherence is possible in principle if the dissipation in the waveguide cavity is also reduced. In practice, this will require an advance in integrated waveguide technology to achieve smaller propagation losses than the current state of the art.

\section{More advanced experiments based on fundamental capabilities}

The experiments that we have studied in some detail in the previous
subsection serve to demonstrate both the ability to couple separate cavities
and to create effective interactions between the atoms that are held
in each cavity. These are essential capabilities that provide the basis
for a wide variety of more challenging experiments for which we would
like to briefly outline some examples here.

\subsection{Preparation of entangled states by propagation}

A stringent test of the coherence properties of transport in extended chains
of cavities is provided by the transport of quantum entanglement through the
chain. Here one might envisage the initial creation of entanglement by controlled
quantum dynamics either between directly neighbouring cavities to generate
entangled two-mode squeezed states between two cavity modes or between atomic
qubits held in different resonators. Most promising in this context though
appears to be the generation of entanglement between an atomic qubit and the
photonic degree of freedom of the cavity it is residing in. Once created, the
coherent coupling between constituents of the cavity array will lead to transport
of the state of the photonic degree of freedom as described in a harmonic chain
\cite{PlenioHE04}. At a suitably chosen time when
we expect maximal entanglement between the atomic qubit and the photonic
degree of freedom of a distant cavity we would then probe for the entanglement
either by quantum state tomography or by measurement of restricted sets of
observables as described in \cite{AudenaertP06}.

\subsection{Probing coherence by transport measurements}

Another approach to determining coherence in cavity arrays exploits the
quantum coherent effect of dynamical localization \cite{DunlapK86,HolthausH96}
in a harmonic chain \cite{VaziriP10} governed by a Hamiltonian of the type
\begin{equation}
        H = \sum_{k=1}^N \hbar(\Omega_0 + \Omega_1\cos\omega t)k
        \sigma_k^{+}\sigma_k^{-} +
        \sum_{k=1}^{N-1} \hbar c (\sigma_k^{+}\sigma_{k+1}^{-}
        + \sigma_k^{-}\sigma_{k+1}^{+}),
\end{equation}
that is, a chain of sites subject to a hopping interaction with strength $c$
and a static and a time-varying field that affect their on-site energies
energies via $\Omega_0$ and $\Omega_1$ respectively. This may be realised
by a spin model or, alternatively, by a set of harmonic oscillators replacing
$\sigma^{+}$ ($\sigma^{-}$) by $a^{\dagger}$ ($a$) respectively. In a
spin model the variation of the on-site energies could be obtained by
a time dependent magnetic field gradient, which induces a shift of the energy
levels. For harmonic oscillator systems the cavity resonance frequencies
would have to be shifted for example via modulations of the resonator lengths
or of the refractive index inside a resonator due to the presence of a detuned
two level system.

The existence of the dynamical localization and its coherent character may
be seen by moving to an interaction picture defined by $|\tilde{\psi}(t)\rangle = e^{-i A(t) \sum_k k\sigma_k^{+}\sigma_k^{-}}|\psi(t)\rangle$ with $A(t) = -\Omega_0 t -
(\Omega_1/\omega) \sin(\omega t)$
in which the time-dependent on-site energies vanish at the expense of introducing
a time-dependent coupling strengths between the neighbouring sites \cite{HolthausH96}.
For small coupling strength $c$ the hopping dynamics is negligible over the interval $[-\pi/\omega,\pi/\omega]$ and then we find the effective time-averaged
Hamiltonian
\begin{equation}
        \label{effective}
        H_I = \sum_{k=1}^{N-1} \hbar c
        J_{\Omega_0/\omega}\left(\Omega_1/\omega\right)
        (\sigma_k^{+}\sigma_{k+1}^{-} + \sigma_k^{-}\sigma_{k+1}^{+})\,.
\end{equation}
If $\Omega_0/\omega=n$ and if $\Omega_1/\omega$ coincides with a zero of the Bessel function $J_n$, then the evolution of a wave-packet becomes
periodic in time, the spreading of the wave-packet is suppressed, and hence transport is suppressed. This expresses itself as resonances in the
ability of the chain to transport excitation, which one can
observe directly without the need for process tomography.
In the averaging that gives rise to the Bessel function, it is the destructive interference of amplitudes that suppresses the transport. In a classical system, this interference is absent and there is no localisation. The strength of dynamic localization can in fact be used to infer the level of quantum
coherence in the system without needing detailed measurements or tomography,
but depending purely on the quality of the transport \cite{VaziriP10}.

\subsection{Long distance entanglement}

Systems with finite correlation length, such as the 1-D Heisenberg and
XX models, allow sizeable ground-state end-to-end entanglement,
independent of the size of the system, provided that simple patterns
of site-dependent couplings are selected. This phenomenon has been
termed Long-Distance Entanglement (LDE) and can be implemented in cavity
arrays \cite{GiampaoloI09,GiampaoloI10}. The realisation of a model
with spin-spin interactions of XX type requires local control, as described in section 6, to create position-dependent coupling between neighbouring sites. In particular it will be necessary to couple the
first and last sites of the chain to the bulk of the chain weakly in order
to obtain strong entanglement between those two sites in the ground state of
the chain. A chain of $4$ cavities suffices to demonstrate this effect.

\subsection{Adiabatic quantum state preparation}

The generation of complex quantum states may be achieved by a sequence of
elementary quantum operations correlating pairs of sub-systems. However, under
realistic conditions the resulting fidelity will tend to scale
very badly with increasing number of particles and complexity of the quantum state.
It may therefore be preferable to take advantage of the natural dynamics. For example, we might first prepare the system in a product state without interactions between sites, then slowly switch on the coupling between neighbouring sites. As a consequence of the adiabatic theorem, the system will remain in its
instantaneous ground state, which thanks to the interactions becomes
a complex entangled state. This highly entangled state of the atomic
degrees of freedom could then serve as a deterministic source of highly entangled
photons. Indeed, the atomic degrees of freedom may be mapped onto the photonic mode at any desired time and photons allowed to escape into fibres.

\subsection{Non-adiabatic creation of entanglement}

The generation of
entanglement between distant sites in the cavity array is complicated
by the fact that interactions are between nearest neighbours.
Distant particles may be entangled through a sequence of
nearest-neighbour interactions, but the efficiency of this preparation
decreases very rapidly with distance. An alternative is to use the natural dynamics
of states that are not eigenstates of the system Hamiltonian.
Thanks to the availability of local control in a cavity array, it is possible for example
to prepare the atoms inside the cavities in a product state.
For most spin Hamiltonians, such as those generated in sec 6, this initial
state will evolve in time to become highly entangled. In
particular, significant two-particle entanglement will build up between
sites whose distance is proportional to twice the speed of sound multiplied
with the waiting time \cite{EisertPHB04}. A basic demonstration
of such an experiment is already possible with three sites but
may be scaled to larger arrays as the strength of the entanglement
decreases only slowly with distance.

\section{Summary and Outlook \label{sec:outlook}}

In summary, we have presented a practical way to realise an array of coupled cavities that could be used for quantum simulation. The device consists of open Fabry-Perot microcavities that are coupled via a waveguide chip. We have demonstrated that, under suitable conditions, the dynamical evolution is well approximated by a Jaynes-Cummings lattice
Hamiltonian and that it is suitable for implementing effective spin Hamiltonians. We have outlined experiments, both basic and more advanced, that could be carried out in such a device to achieve controlled quantum dynamics.

\subsection*{Acknowledgements \label{subsec:acknowledgements}}

We gratefully acknowledge financial support from the European Union (HIP), the Austrian Nano-initiative (PLATON-NAP) the FWF, the UK EPSRC and Germany's Alexander von Humboldt Foundation. MJH acknowledges financial support from the German Research Foundation (DFG) via the Emmy Noether project HA 5593/1-1 and the SFB 631. EAH was supported by the Royal Society.

\appendix

\section{Derivation of effective spin interactions \label{sec:adia_elim}}

The Schr\"odinger equation containing the Hamitonian $H$ as in equation (\ref{eq:ham-lambda-chain-ip}) reads,
\begin{equation}
\frac{d}{dt} |\Psi,t \ket = - i H(t) |\Psi,t \ket .
\end{equation}
This equation can formally be integrated to yield
\begin{equation}
|\Psi,t + T \ket = |\Psi,t\ket - i \int_{t}^{t+T} ds H(s) |\Psi,s \ket \, .
\end{equation}
Iterating the right hand side results in,
\begin{eqnarray} \label{eq:dyson-iterate}
& & |\Psi,t + T \ket = |\Psi,t \ket + \\
& + & \sum_{n=1}^{\infty} (- i)^{n} \int_t^{t+T} dt_{n} H(t_{n}) \int_{t}^{t_{n}} dt_{n-1} H(t_{n-1}) \dots
\int_{t}^{t_{1}} dt_{0} H(t_{0})|\Psi,t \ket \, . \nonumber
\end{eqnarray}
An effective Hamiltonian that accurately describes processes which happen on a time scale $T$ can now be found by performing the time integrations
and identifying the dominant terms.
For our derivation, we assume that at time $t$, the excited states of the atoms are not occupied and that no photons are present,
\begin{equation}
\bra e | \Psi,t \ket = 0 \quad \mathrm{and} \quad  a_{k} | \Psi,t \ket = 0.
\end{equation}
Furthermore, we assume for the parameters in $H(t)$
\begin{eqnarray} \label{eq:adiaelimcond}
\left|\frac{\Omega_{x}}{2 \Delta_{x}}\right| \ll 1 \quad & \mathrm{for} & \quad x = a,b \\
\left|\frac{\Omega_{x} g_{y,j,k}}{2 \Delta_{x}(\Delta_{x} - \delta_{k}^{y})}\right| \ll 1 \quad & \mathrm{for} & \quad x,y = a,b
\end{eqnarray}
We keep terms up to $n = 3$ on the right hand side of equation (\ref{eq:dyson-iterate}) and, by virtue of equation (\ref{eq:adiaelimcond}), neglect all oscillating terms to arrive at
\begin{equation}
|\Psi,t + T \ket = \left(1 + \sum_{\mu=1}^{4}\frac{\left(- i H_{0} T\right)^{\mu}}{\mu!} + \sum_{\nu=1}^{2}\frac{\left(- i H_{1} T\right)^{\nu}}{\nu!} - i H_{2} T \right) |\Psi,t\ket \, ,
\end{equation}
where
\begin{eqnarray*}
H_{0} & = & \delta \sum_{j=1}^{N} |b_{j}\ket \bra b_{j}|, \\
H_{1} & = & - \frac{|\Omega_{b}|^{2}}{4 \Delta_{b}} \sum_{j=1}^{N} |b_{j}\ket \bra b_{j}| - \frac{|\Omega_{a}|^{2}}{4 \Delta_{a}} \sum_{j=1}^{N} |a_{j}\ket \bra a_{j}| \quad \mathrm{and}\nonumber \\
H_{2} & = & \sum_{j=1}^{N} \frac{|\Omega_{b}|^{2}}{4 \Delta_{b}^{2}} \left[ \frac{|\Omega_{b}|^{2}}{2 \Delta_{b}} + \frac{|\Omega_{a}|^{2}}{4 (\Delta_{a} - \Delta_{b})} + \sum_{k} \left( \frac{|g_{b,j,k}|^{2}}{\delta_{k}^{b} - \Delta_{b}} + \frac{|g_{a,j,k}|^{2}}{\delta_{k}^{a} - \Delta_{b}} \right) \right] |b_{j}\ket \bra b_{j}| \nonumber \\
& + & \sum_{j=1}^{N} \frac{|\Omega_{a}|^{2}}{4 \Delta_{a}^{2}} \left[ \frac{|\Omega_{a}|^{2}}{2 \Delta_{a}} + \frac{|\Omega_{b}|^{2}}{4 (\Delta_{b} - \Delta_{a})} + \sum_{k} \left( \frac{|g_{a,j,k}|^{2}}{\delta_{k}^{a} - \Delta_{a}} + \frac{|g_{b,j,k}|^{2}}{\delta_{k}^{b} - \Delta_{a}} \right) \right] |a_{j}\ket \bra a_{j}| \nonumber \\
& + & \sum_{j \neq l} \left( \frac{\Omega_{a} \Omega_{b}^{\star}}{4 \Delta_{a} \Delta_{b}} \sum_{k} \frac{g_{b,j,k}^{\star} g_{a,l,k}}{\delta_{k}^{b} - \Delta_{a}} \, |b_{j}\ket \bra a_{j}| \otimes |b_{l}\ket \bra a_{l}| + \mathrm{H.c.} \right) \nonumber \\
& + & \sum_{j \neq l} \left( \frac{|\Omega_{b}|^{2}}{4 \Delta_{b}^{2}} \sum_{k} \frac{g_{a,j,k}^{\star} g_{a,l,k}}{\delta_{k}^{a} - \Delta_{b}} \, |b_{j}\ket \bra a_{j}| \otimes |a_{l}\ket \bra b_{l}| + \mathrm{H.c.} \right) \nonumber
\end{eqnarray*}
Provided we can choose the time scale $T$ such that $||H_{0}T|| \ll 1$, $||H_{1}T|| \ll 1$ and $||H_{2}T|| \ll 1$ but
$\Delta_{x} T \gg 1$ and $\delta_{k}^{x} T \gg 1$ ($x = a,b$), we can write an effective Schr\"odinger equation,
\begin{equation}
\frac{|\Psi,t + T \ket - |\Psi,t\ket}{T} \approx -i \left( H_{0} + H_{1} + H_{2} \right) |\Psi,t\ket.
\end{equation}
On time scales $T$, the dynamics of our system is thus accurately described by the effective Hamiltonian $\mathcal{H} = H_{0} + H_{1} + H_{2}$,
which is identical to the effective spin Hamitonian of equation (\ref{eq:spinham}) up to an irrelevant global constant.

\section{Explicit expressions for the spin parameters in the $N=2$ case \label{sec:N2exprs}}
\begin{eqnarray}
B|_{N=2} & = & \frac{\delta}{2} - \frac{|\Omega_{b}|^{2}}{8 \Delta_{b}^{2}} \left[ \Delta_{b} - \frac{|\Omega_{b}|^{2}}{2 \Delta_{b}} - \frac{|\Omega_{a}|^{2}}{4 (\Delta_{a} - \Delta_{b})} \right. \\
& & \left. - \frac{|g_{b}|^{2} (\delta_{b} - \Delta_{b})}{(\delta_{b} - \Delta_{b})^{2} - J^{2}} - \frac{|g_{a}|^{2} (\delta_{a} - \Delta_{b})}{(\delta_{a} - \Delta_{b})^{2} - J^{2}}\right] \nn \\
& + & \frac{|\Omega_{a}|^{2}}{8 \Delta_{a}^{2}} \left[ \Delta_{a} - \frac{|\Omega_{a}|^{2}}{2 \Delta_{a}} - \frac{|\Omega_{b}|^{2}}{4 (\Delta_{b} - \Delta_{a})} \right. \nn \\
& & \left. - \frac{|g_{a}|^{2} (\delta_{a} - \Delta_{a})}{(\delta_{a} - \Delta_{a})^{2} - J^{2}} - \frac{|g_{b}|^{2} (\delta_{b} - \Delta_{a})}{(\delta_{b} - \Delta_{a})^{2} - J^{2}} \right], \nn \\
J_{j,l}|_{N=2} & = & \frac{\Omega_{a} \Omega_{b}^{\star}}{4 \Delta_{a} \Delta_{b}} \frac{g_{b}^{\star} g_{a} J}{(\delta_{b} - \Delta_{a})^{2} - J^{2}}, \\
K_{j,l}|_{N=2} & = & \frac{|\Omega_{b}|^{2}}{4 \Delta_{b}^{2}} \frac{|g_{a}|^{2} J}{(\delta_{a} - \Delta_{b})^{2} - J^{2}} +
\frac{|\Omega_{a}|^{2}}{4 \Delta_{a}^{2}} \frac{|g_{b}|^{2} J}{(\delta_{b} - \Delta_{a})^{2} - J^{2}}.
\end{eqnarray}
Here, $\delta_{a} = \omega_{e} - \omega_{C}$ and $\delta_{b} = \omega_{e} - (\omega_{b} - \delta) - \omega_{C}$.

\section{Master equation for the dynamics of two three-level atoms \label{sec:full_lambda}}
The three levels of $^{87}$Rb depicted in figure \ref{lambdalevel} are two of the $5s^{2}S_{1/2}$ ground-states,   $|a\ket = |F=1,m_{F}=1\ket$, $|b\ket = |F=2,m_{F}=1\ket$, and the $5p^{2}P_{3/2}$ excited state  $|e\ket = |F'=2,m_{F'}=2\ket$. In order to compute the dynamical evolution of this system in the presence of $\sigma^{+}$ light, one needs to include spontaneous emission on the $\pi$ transition to the $5s^{2}S_{1/2}$ ground state  $|x\ket=|F=2,m_{F}=2\ket$, which is outside the three-level system under consideration. The dynamics is thus described by the master equation,

\begin{eqnarray}
\frac{d\rho}{dt} = -i \left[H,\rho\right] &+ & \frac{\gamma}{2} \sum_{j} ( 2 \sigma_{j}^{ae} \rho \sigma_{j}^{ea} - \sigma_{j}^{ee} \rho - \rho \sigma_{j}^{ee})\\
 &+ & \frac{\gamma}{6} \sum_{j} ( 2 \sigma_{j}^{be} \rho \sigma_{j}^{eb} - \sigma_{j}^{ee} \rho - \rho \sigma_{j}^{ee})\nn\\
 &+ & \frac{\gamma}{3} \sum_{j} ( 2 \sigma_{j}^{xe} \rho \sigma_{j}^{ex} - \sigma_{j}^{ee} \rho - \rho \sigma_{j}^{ee})\nn\\
 & + & \tilde{\kappa} \sum_{j} ( 2 a_{j} \rho a_{j}^{\dagger} - a_{j}^{\dagger} a_{j} \rho - \rho a_{j}^{\dagger}\nn a_{j}),
\end{eqnarray}
where $\sigma_{j}^{\alpha \beta} = |\alpha_{j}\ket \bra \beta_{j}|$ and the atomic transition rates are weighted according to the squares of the respective dipole matrix elements .


\end{document}